\documentclass[showpacs, preprintnumbers, nofootinbib, aps, prd,
superscriptaddress,10pt, showkeys, notitlepage,twocolumn]{revtex4-1}

\usepackage[linktocpage,breaklinks]{hyperref}
\usepackage[usenames,dvipsnames]{xcolor}

\usepackage{amsmath}
\usepackage{amssymb}
\usepackage{amsfonts}
\usepackage{txfonts}
\usepackage{bm}
\usepackage{stmaryrd}
\usepackage{tensor}
\usepackage{mathrsfs}
\usepackage[utf8]{inputenc}
\usepackage{url}

\usepackage{graphicx}
\usepackage{epsfig}
\usepackage{epstopdf}
\usepackage[normalem]{ulem}

\usepackage{natbib}
\usepackage{hyperref}
\usepackage{cleveref}
\hypersetup{colorlinks=true,
            citecolor=Black,
            linkcolor=Black,
            urlcolor=Black}

\newcommand{\dd}{{\rm d}}

\newcommand{\ii}{{\rm i}}
\newcommand{\calM}{{\cal M}}

\newcommand{\EdGB}{{\mbox{\tiny EdGB}}}
\newcommand{\dCS}{{\mbox{\tiny dCS}}}

\begin{document}

\title{Fundamental Physics Implications on Higher-Curvature Theories
 	\\ from the Binary Black Hole Signals in the LIGO-Virgo Catalog GWTC-1}

\begin{abstract}
Gravitational-wave astronomy offers not only new vistas into the realm of
astrophysics, but it also opens an avenue for probing, for the first time, general
relativity in its strong-field, nonlinear, and dynamical regime, where the
theory's predictions manifest themselves in their full glory.
We present a study of whether the gravitational-wave events detected so far by
the LIGO-Virgo scientific collaborations can be used to probe higher-curvature
corrections to general relativity.
In particular, we focus on two examples: Einstein-dilaton-Gauss-Bonnet
and dynamical Chern-Simons gravity.
We find that the two events with a low-mass $m \approx 7 M_{\odot}$ BH
(GW151226 and GW170608) place stringent constraints
on Einstein-dilaton-Gauss-Bonnet gravity, ${\alpha}^{1/2}_{\EdGB} \lesssim 5.6$ km,
whereas dynamical Chern-Simons gravity remains unconstrained by
the gravitational-wave observations analyzed.
\end{abstract}

\author{Remya Nair}
\affiliation{eXtreme Gravity Institute,
Department of Physics, Montana State University, Bozeman, Montana 59717 USA}

\author{Scott Perkins}
\affiliation{eXtreme Gravity Institute,
Department of Physics, Montana State University, Bozeman, Montana 59717 USA}
\affiliation{Department of Physics, University of Illinois at Urbana-Champaign, Urbana, Illinois 61801, USA}

\author{Hector O. Silva}
\affiliation{eXtreme Gravity Institute,
Department of Physics, Montana State University, Bozeman, Montana 59717 USA}
\affiliation{Department of Physics, University of Illinois at Urbana-Champaign, Urbana, Illinois 61801, USA}

\author{Nicol\'as Yunes}
\affiliation{eXtreme Gravity Institute,
Department of Physics, Montana State University, Bozeman, Montana 59717 USA}
\affiliation{Department of Physics, University of Illinois at Urbana-Champaign, Urbana, Illinois 61801, USA}

\date{\today}

\maketitle
\emph{Introduction.--}~General relativity (GR) remains our most accurate theory
for the gravitational interaction~\cite{Will:2014kxa}. The centennial theory
has passed a plethora of tests ranging from those carried out in the
weak-gravitational field and low-velocity regime of our Solar System, to those
performed in the extreme, nonlinear, and highly dynamical regime of plunging
and merging compact objects, such as neutron stars (NSs) and black holes
(BHs)~\cite{Yunes:2009ke,LIGOScientific:2019fpa}.
The agreement between the observations and predictions is dazzling. In turn,
any new observation that may hint toward a failure of GR will require us to
revisit its foundations.
Experimental tests of GR not only allow us to place its foundational principles
on solid  ground, but they also allow us to constrain (or even rule out)
contending theories that violate one or more of its pillars.
Such contending theories have been developed to address certain outstanding
mysteries in recent observations~\cite{Clifton:2011jh,Berti:2015itd}, such as
the enigmatic late-time acceleration of the
Universe~\cite{Riess:1998cb,Perlmutter:1998np}, the matter-antimatter asymmetry
in our Universe~\cite{Spergel:2003cb,Canetti:2012zc}, and the rotation curve of
galaxies~\cite{Sofue:2000jx,Bertone:2016nfn}.

One broad class of modifications to GR that arise naturally in attempts
to unify gravity with quantum mechanics are quadratic gravity theories~\cite{Yagi:2015oca}.
This class of theories is characterized by the presence of an additional
scalar degree of freedom (violating the GR pillar that gravity is
mediated by a single metric tensor) coupled to a higher-order curvature scalar.
Two preeminent examples of such theories are Einstein-dilaton-Gauss-Bonnet (EdGB)
and dynamical Chern-Simons (dCS) gravity~\cite{Alexander:2009tp}.
Both of these emerge naturally in the context of grand unified theories (string
theory in particular) in the low-energy limit upon dimensional reduction.
Phenomenologically, they predict BHs that carry a nontrivial scalar field,
resulting in a violation of the strong equivalence principle.

Aside from these theoretical motivations, are EdGB and dCS gravity consistent
with experimental tests?
Within the confines of our Solar System, the parameterized-post-Newtonian
parameters of EdGB gravity are identical to those of GR~\cite{Sotiriou:2006pq},
and therefore the theory survives all experimental tests in this regime.
In contrast, dCS gravity contains a nonzero (different from GR) parameter
that leads to modifications in the Lense-Thirring precession of spinning
bodies~\cite{Alexander:2007zg,Alexander:2007vt}. Solar System
experiments such as LAGEOS~\cite{Ciufolini:2004} and Gravity Probe B~\cite{Everitt:2011hp}
can place constraints on the dCS coupling parameter, but due to the weak curvatures in
the Solar System, these constraints are extremely weak~\cite{AliHaimoud:2011fw}.
Exquisitely accurate binary-pulsar observations suffer the same fate.
The post-Keplerian motion of NS binaries in EdGB and dCS gravity is
very similar to that in GR, because the scalar field sourced by such
stars is suppressed relative to that created by BHs, which means that
constraints with present day binary pulsar observations are not
possible~\cite{Yagi:2013mbt,Yagi:2015oca}.

This leaves us with gravitational wave (GW) observations as a last
resort. In recent years, considerable effort has been made in modeling the
inspiral~\cite{Yagi:2011xp,Loutrel:2018rxs,Loutrel:2018ydv},
merger~\cite{Okounkova:2017yby,Okounkova:2018abo,Okounkova:2018pql,Witek:2018dmd}
and ringdown~\cite{Cardoso:2009pk,Cardoso:2009pk,Blazquez-Salcedo:2016enn}
phases of compact binaries in these two theories. One could then imagine comparing
such waveform models against the GW data to determine how small the EdGB and dCS
coupling parameters must be in order to be consistent with statistical noise.
We build on these efforts and use the constraints on GR deviations
obtained by the LIGO-Virgo collaboration (LVC)~\cite{LIGOScientific:2018mvr}
to analyze whether these two theories can be constrained with the binary BH
events detected during the first two observation runs of the LVC.
More specifically, we will consider the binary BH events in the LIGO-Virgo Catalog GWTC-1
GW150914~\cite{Abbott:2016blz,TheLIGOScientific:2016qqj},
GW151226~\cite{Abbott:2016nmj},
GW170104~\cite{Abbott:2017vtc},
GW170608~\cite{Abbott:2017gyy} and
GW170814~\cite{Abbott:2017oio}
for which the posteriors on theory-independent GR modifications,
obtained through a Markov-chain Monte-Carlo (MCMC) exploration of
the parameter space, have been made public~\cite{LIGOScientific:2019fpa,LIGOposteriors}.

\emph{Quadratic gravity.}~dCS and (decoupled) EdGB
theories are defined in vacuum by the Lagrangian density~\cite{Yagi:2015oca}
%
\begin{align}
\mathscr{L}_{\dCS} \! &=
\kappa\, R
- \frac{1}{2} \nabla_{\mu}\vartheta_{\dCS}\nabla^{\mu}\vartheta_{\dCS}
+ \frac{\alpha_{\dCS}}{4}\, \vartheta_{\dCS}\, {}^{\ast}RR,
\\
\mathscr{L}_{\EdGB} \! &=
\kappa\, R
- \frac{1}{2} \nabla_{\mu}\vartheta_{\EdGB}\nabla^{\mu}\vartheta_{\EdGB}
+ \alpha_{\EdGB} \, \vartheta_{\EdGB} \, \mathscr{G},
\end{align}
where $\kappa \equiv (16 \pi)^{-1}$, $g$ is the determinant
of the metric $g_{\mu\nu}$,
${}^{\ast}RR =R_{\nu\mu\rho\sigma} {}^{\ast}R^{\mu\nu\rho\sigma}$
is the Pontryagin density (constructed in terms of the Riemann tensor and
its dual),
$\mathscr{G} = R^2 - 4 R_{\mu\nu}R^{\mu\nu} + R_{\mu\nu\rho\sigma}R^{\mu\nu\rho\sigma}$
is the Gauss-Bonnet density (where $R$ and $R_{\mu\nu}$ are the Ricci scalar
and tensor), and  we have used geometric units, in which $c = 1 = G$.
These quadratic-in-curvature scalars are coupled to a massless scalar
(pseudo-scalar) field $\vartheta_{\EdGB}$ ($\vartheta_{\dCS}$) through the
coupling constants $\alpha_{\EdGB}$ ($\alpha_{\dCS}$), with units of
$({\rm{length}})^{2}$.
In EdGB, the coupling to the Gauss-Bonnet density is usually of exponential
form. We here work in the decoupling (effective field theory) limit, in which
the exponential is expanded to linear order~\cite{Yagi:2015oca}.

To ensure the perturbative well-posedness of these theories, we work in the small-coupling
approximation, in which modifications to GR are \textit{small deformations}.
This is a justified assumption given the agreement of GR with various observations,
GW events included.
It is convenient to define the dimensionless parameter
$\zeta_{\dCS,\EdGB} \equiv \alpha^{2}_{\dCS,\EdGB} / (\kappa \, {\ell}^4)$,
where ${\ell}$ is the typical mass scale of a system.
For the small-coupling approximation to be valid we must have $\zeta_{\dCS,\EdGB}<1$ or ${\alpha}_{\dCS, \EdGB}^{1/2} / m_s \lesssim 0.5$ where $m_s$ is the smallest
mass scale involved in the problem. Note that $0.5$ is a rough threshold which we use as a proxy for the validity of the approximation.

Consistency with Solar System experiments (in dCS) and with low-mass x-ray binary
observations (in EdGB) impose the
upper bounds $\alpha^{1/2}_{\dCS} \leq {\cal O}\, (10^{8} \; {\rm{km}})$~\cite{AliHaimoud:2011fw,Alexander:2009tp} and
$\alpha^{1/2}_{\EdGB} \leq {\cal O}\, (2 \; {\rm{km}})$~\cite{Yagi:2012gp}.

How can the GWs emitted by BH binaries in these theories be different from GR's
predictions?
In both theories, BHs support a nontrivial scalar field -- dipolar in
dCS~\cite{Yunes:2009hc} and monopolar in EdGB~\cite{Kanti:1995vq} -- which
results in the emission of scalar quadrupole (in dCS) and scalar dipole (in
EdGB) radiation during the inspiral. This additional channel for binding energy
loss results in modification to the GW phase, which appear at 2PN [In
the PN formalism, quantities of interest such as the conserved energy, flux
etc. can be written as expansions in $(v/c)$, where $v$ is the characteristic
speed of the binary system and $c$ is the speed of light. ${\cal{O}}((v/c)^n)$
corrections counting from the Newtonian (leading order GR) term are referred to
as $(n/2)$PN-order terms ~\cite{Blanchet:2013haa,Damour:2016bks}.] (for dCS)
and -1PN (for EdGB) order.
In dCS gravity, the scalar field also introduces a quadrupolar correction to
the binary BH spacetime, introducing 2PN corrections to the binding energy,
which in turn affect the GW phase evolution at the same PN order.
Hereafter, we use these facts, together with the estimates of the GW model
parameters and the posterior distributions released
in~\cite{LIGOScientific:2018mvr,LIGOScientific:2019fpa}, to investigate how
well (if at all) the observed GW events in the LVC catalog can be used to
constrain these theories.

\emph{Order of magnitude constraints.--}~It is illuminating to start with a
simple order-of-magnitude calculation to assess if the binary BH events
detected by LIGO-Virgo can place any constraints on dCS and EdGB gravity.
Consider the Fourier domain gravitational waveform $\tilde{h} = A(f) \exp[\ii
\Psi(f)]$, and for simplicity we assume that the spins of the compact objects
are (anti)aligned to the orbital angular momentum. Under these assumptions, the
leading-order modification to the Fourier phase $\Psi(f)$ takes on the
parametrized post-Einsteinian (ppE) form~\cite{Yunes:2009ke} $\Psi = \Psi_{\rm
GR} + \beta \, (\pi \calM f)^{b}$, where $b_{\dCS}=-1/3$ in dCS gravity (a 2PN
correction) and $b_{\EdGB} = -7/3$ in EdGB gravity (a -1PN correction). The
amplitude coefficient $\beta$ is
\begin{align}
\beta_{\dCS} &= - \frac{5}{8192} \frac{\zeta_{\dCS}}{\eta^{14/5}}
\frac{(m_1\, s_2^{\dCS} - m_2\, s_1^{\dCS})^2}{m^2}
\nonumber \\
&\quad + \frac{15075}{114688} \frac{\zeta_{\dCS}}{\eta^{14/5}} \frac{1}{m^2}
\left(
m_2^2 \, \chi_1^2 - \frac{350}{201} m_1 \,m_2 \,\chi_1 \,\chi_2 + m_1^2 \, \chi_2^2
\right)
\label{eq:beta_dcs}
\end{align}
in dCS gravity\footnote{
Our expression for $\beta_{\dCS}$ is different from that presented, e.g.
in~\cite{Yunes:2016jcc,Tahura:2018zuq}.
First, we corrected an error in the rate of scalar radiation emission
$\dd \delta {E}^{(\vartheta)} / \dd t$, which propagates to the final expression
for $\beta_{\dCS}$~\cite{Yagi:2013mbt}.
Second, we do not expand the charge $s_i^{\dCS}$ to leading order in $\chi_i$ as
has been done in the past.
The reason is the following: the binding energy contribution to $\beta_{\dCS}$ in Eq.~\eqref{eq:beta_dcs}
only contains the quadrupole moment to $O(\chi_i^{2})$.
In principle, there will be a $O(\chi_{i}^{4})$ correction to it,
which will also enter at 2PN order and has not been calculated yet.
Thus, unlike in the EdGB case, we cannot calculate the
dCS correction at 2PN order to also all orders in the spins.
To estimate how robust our bounds are to the absence of this quadrupolar contribution,
we include the full expression for $s_{i}^{\dCS}$, in the calculation of $\beta$,
as a proxy for the missing $O(\chi_i^4)$ term.
We checked that all our results are unaffected by using Eq.~\eqref{eq:charge_dcs}
or its leading order in spin expansion.
}
~\cite{Yagi:2012vf} and
\begin{align}
    \beta_{\EdGB} &= - \frac{5}{7168} \frac{\zeta_{\EdGB}}{\eta^{18/5}}
    \frac{\left(m_1^2\, s_2^{\EdGB} - m_2^2\, s_1^{\EdGB} \right)^2}{m^4}\,,
\label{eq:beta_edgb}
\end{align}
in EdGB gravity~\cite{Yagi:2011xp},
where $\calM = (m_1\, m_2)^{3/5} / (m_1 + m_2)^{1/5}$ is the chirp mass,
$\eta = m_1 m_2 / m^2$ (with $m = m_1 + m_2$) is the symmetric mass ratio,
$\chi_{s,a} = (\,\chi_1 \pm \chi_2) / 2$
are the symmetric and antisymmetric dimensionless spin combinations
with $\chi_{i} = \vec{S}_{i} \cdot \hat{L} / m_{i}^2$ the
projections of dimensional spin angular momenta $\vec{S}_{i}$
in the direction of the orbital angular momentum $\hat{L}$ and
\begin{align}
    s_i^{\dCS} &= \frac{2 + 2\,\chi_i^4 - 2 (1-\chi_i^2)^{1/2} - \chi_i^2[3 - 2(1-\chi_i^2)^{1/2}]}{2 \chi_i^3}\,,
    \label{eq:charge_edgb}
    \\
    s_i^{\EdGB} &= \frac{2\, [(1-\chi_i^2)^{1/2} - 1 + \chi_i^2]}{\chi_i^2}\,,
    \label{eq:charge_dcs}
\end{align}
are the dimensionless spin and mass-dependent BH scalar charges,
to all orders in spin, in both theories~\cite{Yagi:2012vf,Yunes:2016jcc,Berti:2018cxi}.
Although $\beta_{\dCS}$ has uncontrolled remainders of ${\cal{O}}(\chi^{4})$, $\beta_{\EdGB}$
is valid to all orders in the spin.
We can obtain an order-of-magnitude bound on $\zeta_{\dCS, \EdGB}$ using the best-fit
parameters from GW170608 and doing a crude Fisher matrix analysis (we use this
particular event as an example because it will allows us to compare our
analytical estimate with more robust calculations later).
Given that the event is consistent with GR, we can ask
how large $\zeta_{\dCS, \EdGB}$ can be and yet remain consistent with the event.
For sufficiently high signal-to-noise ratio (SNR) $\rho$, the accuracy at
which a parameter $\theta^{\,a}$ of the GW model can be estimated from the Cramer-Rao bound~\cite{Finn:1992xs}
$\Delta\theta^{\,a} = \sqrt{(\Gamma^{-1})^{aa}}$
where the Fisher matrix is
\begin{equation}
\Gamma_{ab} \equiv 4\, \textrm{Re}\,\int_{f_{\rm min}}^{f_{\rm max}}
\frac{\partial_{a}\tilde{h}(f)\, \partial_{b}\tilde{h}^{*}(f)}{S_{n}(f)}\, \dd f\,,
\end{equation}
and the asterisk stands for complex conjugation.
The partial derivatives are taken with respect
to the model parameters $\theta^{\,i}$ and $S_{n}(f)$ is the spectral noise density
of the detector. The integration limits denote the lower and upper cutoff
frequencies at which the detector operates. For a rough estimate, it suffices to
neglect correlations between model parameters, and thus, $\Gamma_{ab}$ is
approximately diagonal. With this, one then finds that the variance satisfies
$(\Delta \zeta)^2 = 1 / \Gamma_{\zeta\zeta}$,
which can be evaluated analytically assuming white noise.
This matrix element is dominated by the lower limit of integration $f_{\rm min}$,
and thus, one finds that
\begin{align}
    (\Delta \alpha_{\dCS, \EdGB})^{1/2} &\gtrsim
    \left(1 -  \frac{3b_{\dCS, \EdGB}}{2}\right)^{1/8}
    \frac{(\pi \hat{\calM} f_{\rm min})^{-b_{\dCS, \EdGB}/4}}{(16\pi \hat{\rho})^{1/4}}
 \frac{\hat{m}}{\hat{\beta}^{1/4}_{\dCS, \EdGB}}.
 \nonumber \\
\end{align}
where the overhead hat stands for the best-fit values, with $\zeta_{\dCS,\EdGB}$
set to unity in $\hat{\beta}_{\dCS, \EdGB}$.
As the individual spins $\chi_{i}$ could not be resolved for the events we are
	considering, we assign $\chi_{1} = \chi_{\rm eff} (\, m \, / \, m_1)$ and $\chi_{2} = 0$
	to proceed.
Using $f_{\rm min} = 10$ Hz and the SNR $\hat{\rho}$
and median values for $m_1$, $m_2$ and $\chi_{\rm eff}$, we obtain
$(\Delta \alpha_{\dCS})^{1/2} \approx 28.1$ km and
$(\Delta \alpha_{\EdGB})^{1/2} \approx 1.0$ km at 90\% credibility.
These bounds agrees well with the forecast made
in~\cite{Yagi:2012vf} for dCS and in~\cite{Cornish:2011ys} for EdGB.

\emph{Fisher-estimated constraints on LIGO-Virgo data.--}~We also perform a
fully numerical calculation of the Fisher matrix, by modeling  the binaries
with the phenomenolgical waveform template
\textsc{IMRPhenomD}~\cite{Husa:2015iqa,Khan:2015jqa}
We make similar assumptions for the fiducial parameters as we made to obtain
the order of magnitude constraints and consider 5 GW events, GW150914,
GW151226, GW170104, GW170608, and GW170814 (cf.~Table III
in~\cite{LIGOScientific:2018mvr}).
The bounds obtained for the two most constraining events, GW151226 and GW170608
are shown in Table~\ref{tab:const_sum} and they are in good agreement with our
order-of-magnitude calculation for both theories.
The Fisher-estimated constraints for dCS gravity are not shown because they violate the small coupling approximation, as we will discuss in more detail below.

\begin{table}[t]
\begin{tabular}{l | c c c}
\hline
\hline
System & Method & $\alpha_{\EdGB}^{1/2}$ [km] & $\alpha_{\dCS}^{1/2}$ [km] \\
\hline
Current & Frequentist & $2$ & $10^{8}$ \\
\hline
GW151226 & estimate & 0.9  & 12.6 \\
& Fisher & 6.0 & $-$ \\
& Bayesian & 5.7 & $-$ \\
\hline
GW170608 & estimate & 1.0 & 28.1 \\
& Fisher & 3.9 & $-$ \\
& Bayesian & 5.6 & $-$ \\
\hline
\hline
\end{tabular}
\caption{Current constraints on EdGB and dCS gravity
from low-mass x-ray binary and Solar System observations respectively,
with the Fisher-estimated constraints, and
Bayesian constraints using LVC (testing GR) posteriors for GW151226 and GW170608 ~\cite{LIGOScientific:2019fpa,LIGOposteriors}}.

\label{tab:const_sum}
\end{table}

\begin{figure*}[ht]
  \includegraphics{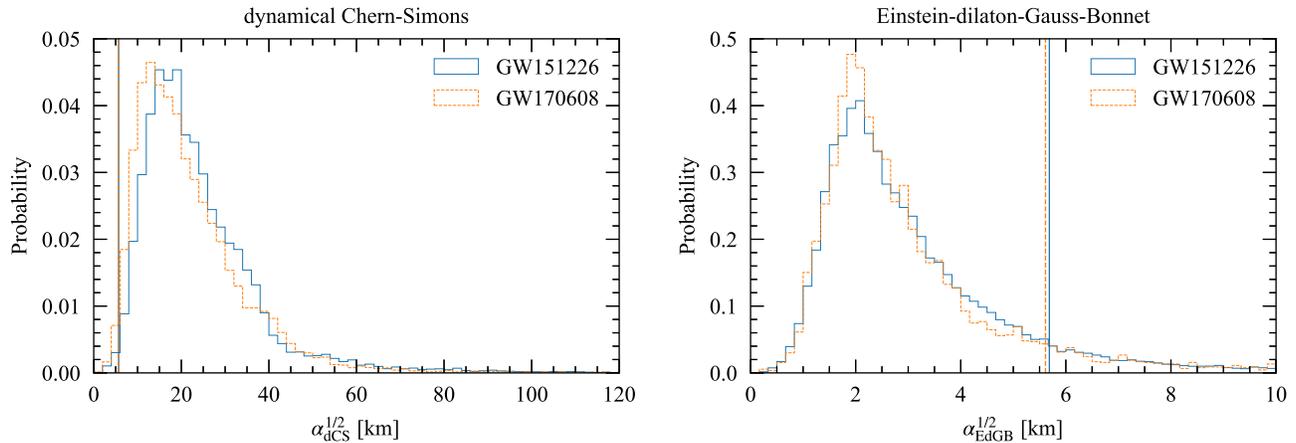}
  \caption{
    Posterior distributions of ${\alpha}_{\dCS}^{1/2}$ (left panel)
    and ${\alpha}_{\EdGB}^{1/2}$ (right panel) obtained using
    GW151226 and GW170608.
    For the GW events shown in both panels, $m_2 / M_{\odot} = 7.7^{+2.2}_{-2.6}$
    (GW151226) and $m_2 / M_{\odot} = 7.6^{+1.3}_{-2.1}$ (GW170104)
    at $90\%$ credibility.
    This implies that the small-coupling approximation
    is valid only when ${\alpha}_{\dCS, \EdGB}^{1/2} \lesssim 5.6$, shown as vertical lines in the plots.
    For dCS gravity (left-panel) we see that most of the support of the
    posterior distributions of these two events
    lays passed the bounds set by the small-coupling approximation.
    Consequently, one cannot place constraints on ${\alpha}_{\dCS}^{1/2}$
    with these two events.
    For EdGB  gravity (right-panel) most ($>90\%$) of the support of posterior
    lays \textit{within} the bound, therefore allowing us to constrain the theory
    with these two events.
    For the other three events, which contain a large $m_2$
    ($\gtrsim 13\, M_{\odot}$) BH~\cite{LIGOScientific:2018mvr}, the vertical lines
    are pushed towards the left, leaving most of the support for the posterior
    outside the small-coupling approximation bound.
    We stress that the location of the peaks in the posteriors
    {\it are not an indication of a deviation from GR}.
    Instead, as detailed in the main text, the lack of support at zero is
    an artifact of the choice of the sampling variable $\delta \phi_i$.
  }
  \label{fig:posteriors}
\end{figure*}
\emph{Bayesian-estimated constraints on LIGO-Virgo data.--}~The LVC recently
released constraints on model-independent deviations from GR to check
consistency of the GW events with GR
predictions~\cite{LIGOScientific:2019fpa,LIGOposteriors}.
The model used to capture these deviations is a variant of
\textsc{IMRPhenomPv2}~\cite{Ajith:2007qp,Ajith:2009bn,Santamaria:2010yb,Husa:2015iqa},
where parameterized relative shifts in the PN coefficients of the Fourier phase
of \textsc{IMRPhenomPv2} are introduced, namely
\begin{equation}
\phi_{i} \rightarrow \phi_{i} \left(1 + \delta \phi_{i}\right)\,,
\end{equation}
with $\delta \phi_{i}$ then treated as additional free parameters in the model.
This modification is nothing but an implementation of the ppE
framework~\cite{Yunes:2009ke,Chatziioannou:2012rf}, as shown explicitly in~\cite{Yunes:2016jcc},
with the mapping
\begin{subequations}
\begin{align}
\label{eq:map1}
\beta_{\dCS} &= \frac{3}{128} \, \phi_{4} \; \delta \phi_{4} \; \eta^{-4/5}\,,
\\
\beta_{\EdGB} &= \frac{3}{128} \; \delta \phi_{-2} \; \eta^{2/5}\,,
\end{align}
\end{subequations}
where $\phi_{4}$ is the GR coefficient of the Fourier phase at 2PN order
(cf.~Appendix B in~\cite{Khan:2015jqa}).
Since the predictions from both dCS and EdGB theories can be mapped to
the ppE framework, one can propagate the LIGO-Virgo bounds on $\delta \phi_{-2}$ and $\delta\phi_{4}$
to constraints on the dCS and EdGB coupling constants.
More specifically, we use the posteriors provided by the LVC
on $\delta \phi_{-2}$ and $\delta \phi_{4}$ to first obtain constraints
on $\beta_{\dCS}$ and $\beta_{\EdGB}$, which we then translate
into constraints on ${\alpha}_{\dCS}^{1/2}$ and ${\alpha}_{\EdGB}^{1/2}$
using Eqs.~\eqref{eq:beta_dcs}-\eqref{eq:beta_edgb}.

The 90\% constraints on ${\alpha}_{\dCS}^{1/2}$ and ${\alpha}_{\EdGB}^{1/2}$
are shown in Table~\ref{tab:const_sum} for the two most constraining events
(GW151226 and GW170608) and the corresponding posterior distributions are shown
in Fig.~\ref{fig:posteriors}. The Fisher estimates, although quite close to the
constraints using posteriors derived from GW data, are overoptimistic since
they assume a Gaussian posterior around the peak, which we see in
Fig.~\ref{fig:posteriors} is not correct. Moreover, since the Fisher analysis
is a point estimate, it is difficult to gauge its robustness.  On the other
hand, a MCMC exploration of the posterior surface helps us evaluate
explicitly how much support the posterior distributions have in the
regions of validity set by the small-coupling approximation.

Constraints on quadratic gravity theories that employ the small-coupling approximation
are robust only provided the former satisfy the requirements of the latter.
For the systems considered, this translates to ${\alpha}_{\dCS, \EdGB}^{1/2} \lesssim 5.6$ km,
which is shown with vertical lines in Fig.~\ref{fig:posteriors}.
For dCS gravity (left panel of Fig.~\ref{fig:posteriors}), more than
$99\%$ of the posterior distribution of ${\alpha}_{\dCS}^{1/2}$ lies
beyond this region of validity for GW151226 and GW170608 and for all the other events we considered.

Consequently, we cannot place constraints on dCS gravity with
the events for which the posteriors samples obtained by LIGO-Virgo have
been released.

For EdGB, the situation is strikingly different. As one can observe in the
right panel of Fig.~\ref{fig:posteriors}, more than $90\%$ of the posterior
distribution falls within the requirements of the small-coupling approximation
for the GW151226 and GW170608 events.
This implies that a 90\% bound of ${\alpha}_{\EdGB}^{1/2} \lesssim 5.6$ km is
statistically meaningful and can be placed on EdGB gravity using these two
events.
This is not the case for the other events (GW150914, GW170104 and GW170814),
for which constraints would violate the small coupling approximation.

We emphasize that the location of the peaks in the posteriors of Fig.~\ref{fig:posteriors}
{\it do not indicate a deviation from GR}.
Rather, the lack of support at zero is an artifact of the choice of the sampling variable
$\delta \phi_i$ and its functional dependence on $\alpha^{1/2}_{\dCS , \EdGB}$.
A uniform prior in $\delta \phi_i$ translates to a nonuniform prior on
$\alpha^{1/2}_{\dCS/ \EdGB}$ with almost no support near $\alpha^{1/2}_{\dCS ,
\EdGB}=0$.
One can reweight the $\alpha^{1/2}_{\dCS/ \EdGB}$ posteriors with the priors to
obtain better estimates, albeit at the cost of introducing binning errors close
to $\alpha^{1/2}_{\dCS/ \EdGB} =0$.

Alternatively, this issue could be avoided by sampling
directly in $\alpha_{\dCS, \EdGB}$ instead of in the generic parameter $\delta \phi_i$.
We expect that this would shift our 90\% bound to the left, thereby
improving our bounds, and hence our constraints
are \emph{conservative and robust} to changes in the sampling variable.

The fact that GW151226 and GW170608 have more constraining power
than their cousins is not surprising. These two events were produced by
binaries in which the secondary BH had the lowest mass ($m_2\approx 7M_\odot$)
of all events in the catalog.
Quadratic gravity theories introduce new length scales, and deviations
from GR are thus proportional to the curvature scale, which for BH binaries
scales inversely with the square of the lowest mass, $m_2^{-2}$.
Hence one can expect the largest deviations for GW151226 and GW170608
and thus, the strongest constraints.
In dCS gravity, the modifications enter at 2PN order, and thus,
they are much more weakly constrained than the EdGB modifications, which enter at -1PN order.
This deterioration in the constraint then implies that a large percentage
of the posterior weight is outside the regime of validity of the small
coupling approximation, rendering the constraint invalid.

\emph{Fundamental physics implications.--}~Our results dramatically constrain
EdGB gravity, essentially confining deviations from GR due to this theory
down to the horizon scale of stellar mass BHs.
These constraints are competitive with those obtained in~\cite{Yagi:2012gp}
(${\alpha}_{\EdGB}^{1/2} \lesssim 2$ km at $95\%$ confidence level)
from the orbital decay on the BH low-mass x-ray binary A0620-00, which probes the theory in a different energy scale.
Our constraints, however, have the advantage of being robust to astrophysical
systematics, unlike those placed in~\cite{Yagi:2012gp} which require assumptions
about the mass transfer efficiency and the specific angular momentum carried by stellar winds.

The constraint we have placed on (decoupled) EdGB gravity is stringent,
limiting this type of quantum-inspired violation of the strong equivalence principle,
the strength of the scalar monopole charge carried by black holes, and the possibility
of using EdGB gravity to explain the late-time acceleration of the universe. However,
our constraints do not directly apply to other functional couplings between the Gauss-Bonnet
density and a scalar field. For example, in models where BHs acquire charges
through spontaneous scalarization~\cite{Silva:2017uqg,Doneva:2017bvd,Silva:2018qhn,Macedo:2019sem,Cunha:2019dwb},
BHs are identical to GR unless they fall within certain mass
intervals (at fixed coupling parameter of the theory) and thereby can (in principle)
mimic binary BH mergers in GR.

Our results also have important implications for restricting
parity violation in the gravitational interaction.
Recently, a broad class of ghost-free,
parity-violating theories, which in four dimensions requires the
presence of a massless scalar field, was presented~\cite{Crisostomi:2017ugk}.
In~\cite{Nishizawa:2018srh,TheLIGOScientific:2017qsa,Monitor:2017mdv}, these theories were tested against the exquisite constraint obtained on the speed of GW propagation from the binary NS event GW170817/GRB 170817A, which estimated that $c_{\rm GW}$ is the same as the speed of light in vacuum to one part in $10^{15}$.
dCS gravity is the only ghost-free, parity-violating theory in four dimensions that is consistent with
this constraint~\cite{Alexander:2017jmt,Nishizawa:2018srh}.
Therefore, our results combined with those by~\cite{Nishizawa:2018srh}, leave
dCS as the single subclass of the broad set of parity-violating
theories of gravity which remains consistent with observations.

Future work could focus on constraints on other modified theories
within the broad class of quadratic gravity models~\cite{Yagi:2015oca}.
Alternatively, one could include GW amplitude corrections due to EdGB and dCS gravity
to determine whether GW constraints become stronger~\cite{Tahura:2018zuq}.
Finally, one could study how well future ground-based and space-based detectors could
constraint quadratic gravity theories, or the type of system that would be ideal to place
constraints the hitherto evasive dCS gravity.

\emph{Acknowledgments.--}~This work was supported by
NASA Grants No.~NNX16AB98G and No.~80NSSC17M0041.
We thank Kent Yagi for discussions and for checking Eq.~\eqref{eq:beta_dcs}.
We thank Alejandro C\'ardenas-Avenda\~{n}o and Katerina Chatziioannou for
helpful discussions and we thank Sandipan Sengupta for useful comments on the
draft.

\bibliographystyle{apsrev4-1}
\bibliography{biblio}

\end{document}